\documentstyle[12pt]{article}
\begin{document}
\baselineskip=24pt
\newcommand{\be}{\beginequation}
\newcommand{\ee}{\endequation}
\newcommand{\cpr}{\mbox{R} \! \!}
\newcommand{\cd}{\mbox{D} \! \!}
\newcommand{\crb}{\bar{\mbox{R}} \! \!}
\title{Coupling of Gravity to Matter\\ via SO(3,2) Gauge Fields \\ 
\vspace{1.2 in} } 
\author{Yuan K. Ha \\ Department of Physics, Temple University \\
Philadelphia, Pennsylvania 19122 U.S.A. \\  \vspace{1.9 in} }
\date{      }
\maketitle
\newpage
\begin{center}
\begin{large}
{\bf Abstract}
\end{large}
\end{center}
We consider gravity from the quantum field theory point of view and introduce
a natural way of coupling gravity to matter by following the gauge principle
for particle interactions. The energy-momentum tensor for the matter fields
is shown to be conserved and follows as a consequence of the dynamics in a
spontaneously broken SO(3,2) gauge theory of gravity. All known interactions 
are described by the gauge principle at the microscopic level.
\newpage
Einstein's theory of gravity is based on simple physical principles, namely,
the principle of equivalence and the principle of general covariance. However,
the matter part of Einstein's equation is quite arbitrary; it is represented only by the energy-momentum tensor of the
matter fields that are present in spacetime. This non-symmetrical situation
is of course a well-known characteristic of the theory. In order to understand
gravity with the other interactions, it would be advantageous to treat all
interactions on the same footing and consider gravity also from the quantum field
theory point of view. We would like to introduce a natural way of coupling
gravity to matter by following a kind of geometrical principle. This is the
same gauge principle which has been so successful in describing the strong,
weak and electromagnetic interactions \cite{1}. The coupling to matter shall
take a rather simple form by the use of a new metric tensor constructed from the
gauge fields in the gravity theory with enlarged gauge symmetry. After
spontaneous symmetry breaking so that the surviving symmetry is the usual
spacetime symmetry, the theory reproduces Einstein's equation with the
correct energy-momentum tensor of the external matter fields from its own 
dynamics.                                                                 

\vspace{.25in}
It is possible to use the gauge principle to arrive at Einstein's theory of
gravity \cite{2}.  Such an approach requires a nonlinear realization of a certain
broken part of the gauge symmetry under consideration.  In order that the gauge
symmetry describes spacetime symmetry, the gauge transformations and the general
coordinate transformations must be interlocked.  This result is achieved through
the vierbein formalism.  The fundamental objects in the theory are the gauge
fields and the action for such a theory can be constructed solely in terms of
the components of the field strength.  The advantage of this approach is that
all known interactions are described in the gauge theory framework at the
microscopic level.  Since general relativity is a macroscopic theory, the use
of the metric tensor $g_{\mu \nu}$, and the Christoffel connection $\Gamma_{\mu
\nu}^{\lambda}$ , may not be suitable at the particle level. We therefore   
avoid the conventional requirement of using the metric as the fundamental 
description in the present approach.

\vspace{.25in}
We consider a gauge theory of gravity based on the anti de Sitter group  SO(3,2)
. Although a gauge theory of gravity was first formulated with the Lorentz group
 SO(3,1)  , the vierbein had to be introduced in an ad hoc way \cite{3}.  An
improvement was subsequently made to consider the Poincare group as the symmetry
group and to interprete
the vierbein, which transform inhomogeneously, as the
gauge fields of translations \cite{4}.  However, as far as local gauge
transformations are concerned, the gauge fields that correspond to Lorentz
transformations and the vierbein are quite independent.  On the other hand, a
de Sitter gauge theory of gravity \cite{5} provides a unifying treatment of both
the Lorentz generators and the generators of translations so that all ten
generators of the Poincare group are initially the generators of the gauge group 
SO(3,2) .  The distinction between the Lorentz generators and the generators of
translations appears later when the anti de Sitter group  SO(3,2)  is
spontaneously broken down to  SO(3,1) .  As a result, those generators in 
SO(3,2)  that are to become the generators of translations are nonlinearly
realized and they play the role of the vierbein in ordinary gravity.  

\vspace{.25in}
For subsequent development, it is necessary to present first the salient features
of the de Sitter gauge theory of gravity.  The ten  SO(3,2)  gauge fields,
$\omega_{\mu}^{AB} = - \: \omega_{\mu}^{BA}$ , where $A , B = 1,2,3,4,5$ are
SO(3,2) indices, are defined on the usual four-dimensional spacetime manifold. 
The action for the theory is constructed from the field strength
\begin{equation}
\mbox{F}_{\mu \nu}^{AB} \equiv \cpr\ _{\mu \nu}^{AB} = \partial_{\mu}
\omega_{\nu}^{AB} - \partial_{\nu} \omega_{\mu}^{AB} \\
 + (\omega_{\mu}^{AC} \omega_{\nu C}^{\mbox{ } \:\:\: B}  -  \omega_{\nu}^{AC}
\omega_{\mu C}^{\mbox{ } \:\:\: B} )  ,
\end{equation}
with the  SO(3,2)  raising and lowering operator $ \; \: \eta^{AB} \; \: = \:
\; \eta_{AB} \: \; = \: \; $ \\  diag $(-1,-1,-1,1,1)$, and is given by
\begin{equation}
\mbox{I}_{\rm G}
= \int d^{4} \! x \: [ \epsilon_{ABCDE} \epsilon^{\mu \nu \rho \sigma}
\cpr\ _{\mu \nu}^{AB} \cpr\ _{\rho \sigma}^{CD} \phi^{E}/m
 +\chi\,( \phi^{A}
\phi_{A} - m^{2} ) ]  .
\end{equation}
The symbol $\epsilon_{ABCDE}$ is the invariant  SO(3,2)  tensor and
$\epsilon^{\mu\nu\rho\sigma}$, the spacetime Levi-Civita symbol.  The fields
$\phi^{A}$ are a set of auxiliary scalar fields with the dimension of a mass and
an  SO(3,2)  five-vector. They are non-propagating but are necessary for
implementing the  SO(3,2)  symmetry of the theory at this point. The full
significance of these fields will be revealed later when matter fields are  
considered. Varying the action with respect to $\chi$ leads to the constraint:  
$\phi^{A}\phi_{A} = - \phi_{1}^{2} - \phi_{2}^{2} - \phi_{3}^{2} + \phi_{4}^{2}
+ \phi_{5}^{2} = m^{2}$.  In order to establish the connection with the usual
four-dimensional theory of gravity, it is necessary to break the  SO(3,2) 
symmetry to that of the Lorentz group SO(3,1)  by choosing a gauge in which
\begin{equation}
\phi_{A} = (0,0,0,0,\phi_{5} = m) .
\end{equation}
Since the constraint does not involve other fields, it is therefore possible
to substitute the condition in Eq.(3) into the action thus eliminating the
fields $\phi_{A}$ . The resulting  SO(3,1)  action then becomes
\begin{equation}
\mbox{I}_{\rm G}
= \int d^{4} \! x \:  \epsilon_{abcd5} \epsilon^{\mu \nu \rho \sigma}
\crb\ _{\mu \nu}^{ab} \crb\ _{\rho \sigma}^{cd}  \: ,
\end{equation}
where $a,b,c,d  = 1,2,3,4$ are now  SO(3,1)  indices, and all the fields in the
effective action transform nonlinearly under  SO(3,2)  but linearly under
SO(3,1).  This type of action is known to be invariant under general coordinate
transformations as well as under local Lorentz transformations. In fact, the
action in Eq.(4) takes exactly the same form as the MacDowell-Mansouri  
formulation \cite{6}, which is equivalent to the standard Hilbert action of
Einstein's gravity, together with a cosmological constant and a Gauss-Bonnet
topological invariant. We may identify the gauge fields $\omega_{\mu}^{a5}$, 
which transform as a natural four-vector in the de Sitter gauge theory, with 
the vierbein $e_{\mu}^{a}$ relating the spacetime and internal indices, i.e.
\begin{equation}
e_{\mu}^{a} = \frac{1}{m}\omega_{\mu}^{a5} .
\end{equation}
For the theory to be physical, it is necessary that  det $e_{\mu}^{a} \neq 0$.
As formulated here pure gravity is an affine theory with no need for
the introduction of a spacetime metric. It is only when matter fields are 
coupled to gravity that it becomes necessary to define the usual metric 
$g_{\mu \nu} = e_{\mu}^{a} e_{a \nu}$, a quantity which is not gauge-invariant
however.

\vspace{.25in}
We are now ready to couple gravity to external matter fields in an  SO(3,2)  
gauge-invariant way. We consider first the case of the complex scalar
field $\varphi$ given by the special relativistic Lagrangian
\begin{equation}
{\rm L}_{\mbox{matter}} = (\rm{D}_{\mu}\varphi)^{\dag} (\rm{D}^{\mu}\varphi)
			   -{\mu}^{2}\varphi^{\dag}\varphi
			   -\lambda(\varphi^{\dag}\varphi)^{2}.
\end{equation}
Here $\rm{D}_{\mu}\varphi$ is a possible covariant derivative of the matter
field that could contain interactions with external gauge fields;
$\mu^2$ and $\lambda$ are arbitrary parameters.
The choice of this Lagrangian is motivated by its occurrence in the
electroweak theory. Thus to the original gravity action in Eq.(2) we would 
include the following matter action,
\begin{equation}
\mbox{I}_{\rm{M}} = \int d^{4} \! x \:
	       \sqrt{-g'}\mbox{ }[g'_{\mu\nu}(\rm{D}^{\mu}\varphi)^{\dag}
			    (\rm{D}^{\nu}\varphi)
			   -{\mu}^{2}\varphi^{\dag}\varphi
			   -\lambda(\varphi^{\dag}\varphi)^{2}],
\end{equation}
where we have introduced a new spacetime metric defined by
\begin{equation}     
g'_{\mu\nu} = {\cal D}_{\mu}\phi^{A}{\cal D}_{\nu}\phi^{B}\eta_{AB}/m^{4},
\end{equation}
with $-g' = \mbox{det} \: g'_{\mu\nu}$. This metric is  SO(3,2)-invariant 
since it is defined in terms of the covariant derivative of the  SO(3,2)  group:
\begin{equation}
{\cal D}_{\mu}\phi^{A} = \partial_{\mu}\phi^{A} 
			+ \omega_{\mu}^{AB}\phi_{C}\eta_{BC},
\end{equation}
and it reduces to the ordinary metric $g_{\mu\nu}$ of general relativity upon
symmetry breaking to  SO(3,1) , i.e. 
\begin{equation}
g'_{\mu\nu} \rightarrow \omega^{a5}_{\mu}\omega^{b5}_{\nu}\eta_{ab}/m^{2}
= g_{\mu\nu},
\end{equation}
where $\eta_{ab} = \mbox{diag}(-1,-1,-1,1)$.
The presence of the factor  $\sqrt{-g'}$
in Eq.(7) ensures that the given action is invariant under general coordinate
transformations as well as under local SO(3,2) transformations.

\vspace{.25in}
It is of special importance to examine the equation of motion of the  SO(3,2) 
scalar fields and remark its significance, even though these fields are
non-propagating. The Euler-Lagrange equation for $\phi^{A}$ ,
obtained by varying the combined action
$\rm{I} = \rm{I}_{\rm{G}} + \rm{I}_{\rm{M}}$ 
in Eq.(2) and Eq.(7), is
\begin{equation}
{\displaystyle
{\cal D}_{\mu}
\left[ \frac{\delta \,{\rm I}}{\delta ({\cal D}_{\mu} \phi^{A}) } \right]
- \frac{\delta \, {\rm I}}{\delta \phi^{A}} = 0  } ,
\end{equation}
or equivalently,
\begin{equation}
{\displaystyle
{\cal D}_{\mu}
\left[ {\cal D}_{\nu} \phi^{A} \frac{\delta \,{\rm I}}{\delta g'_{\mu\nu}}
\right]
- m^{4} \chi \phi^{A} = 0  } .
\end{equation}
With the choice of the gauge condition in Eq.(3), it is found that
\begin{eqnarray}
{\cal D}_{\nu} \phi^{A} & = & m^{2} e^{A}_{\nu} ,\;\;\;\;\;\;A = 1,2,3,4 ;
                      \nonumber     \\
{\cal D}_{\nu} \phi^{5} & = & 0  .
\end{eqnarray}
Equation (12) then takes the following form
\begin{equation}
{\displaystyle
m^{2} {\cal D}_{\mu}
\left[ e^{A}_{\nu} \frac{\delta \, {\rm I}}{\delta g'_{\mu\nu}} \right] = 0 } .
\end{equation}
As we shall find,
this result is completely consistent with general relativity. It is precisely
the statement that the covariant divergence of the energy-momentum tensor
vanishes. In fact, Eq.(14) can be written in the more familiar form of 
Riemannian geometry by identifying the Christoffel symbol through the relation
\begin{equation}
{\displaystyle
e^{A}_{\lambda} \Gamma^{\lambda}_{\mu\nu} = {\cal D}_{\mu} e^{A}_{\nu}  } ,
\end{equation}
and recalling the variational definition of the energy-momentum tensor
in field theory
\cite{7},
\begin{equation}
{\displaystyle
{\rm T}^{\mu\nu} = \frac{2}{\sqrt{-g'}} \frac{\delta \,{\rm I}}{\delta g'_{\mu\nu}} 
 } .
\end{equation}
With the help of the identity $\partial_{\mu}\sqrt{-g'} =
                     \sqrt{-g'} \: \Gamma^{\lambda}_{\lambda\mu}$,
Eq.(14) can be finally obtained as
\begin{eqnarray}
 &  & m^{2} \sqrt{-g'} e^{A}_{\nu}
(\partial_{\mu}\rm{T}^{\mu\nu} + \Gamma^{\nu}_{\mu\lambda}\rm{T}^{\mu\lambda} +
\Gamma^{\lambda}_{\lambda\mu}\rm{T}^{\mu\nu} )  \nonumber  \\
 & = &  m^{2} \sqrt{-g'} e^{A}_{\nu} (\nabla_{\mu}\rm{T}^{\mu\nu} ) = 0    ,
\end{eqnarray}
which is the expression for the conservation of the energy-momentum tensor
as stated.

\vspace{.25in}
When spinors $\psi$ are present in the matter part of the action, the 
fermion kinetic term would be modified according to 
\begin{equation}
i\bar{\psi}\gamma^{\mu}{\rm D}_{\mu}\psi           \rightarrow 
\sqrt{-g'} g'^{\mu\nu} \: i\bar{\psi}e^{A}_{\nu}\Gamma_{A}
( {\rm D}_{\mu} + \frac{1}{4}\omega^{BC}_{\mu}\Gamma_{BC} )\psi .
\end{equation}
Here $\Gamma_{A} = \eta_{AB}\Gamma^{B}$, and $\Gamma^{A} =
(\gamma^{\mu}, \gamma^5)$ are the  SO(3,2) `spin matrices'
obeying the rules:  $\{ \Gamma_{A},\Gamma_{B} \} = 2\eta_{AB}$ , and
$\Gamma_{AB} = \frac{1}{2}[\Gamma_{A} , \Gamma_{B}]$ . The extra term in the
total covariant derivative represents interaction of the fermion with the
gravitational gauge fields and hence a `minimal' coupling of gravity to
the fermion. The fermion mass term would take the form
$\sqrt{-g'}m\bar{\psi}\psi$. If matter gauge fields are further present, an extra
kinetic term of the form
\begin{equation}
{\rm L}_{\mbox{gauge}} = \sqrt{-g'} g'_{\mu\alpha} g'_{\nu\beta}
( -\frac{1}{4} B^{\mu\nu} B^{\alpha\beta} ) ,
\end{equation}
in which $B^{\mu\nu}$ denotes the field strengh, is required. All matter terms
so constructed have local  SO(3,2)  symmetry, each contributing to the
energy-momentum tensor and therefore to the dynamics in Einstein's equation.
From the above development, it is seen that the concept of gauge invariance
is even more fundamental than realized; it is just the right principle for
coupling gravity to matter, whose interactions are already prescribed by the
very same principle. 

\vspace{.25in}
Although Einstein's theory of gravity in its original form is adequate 
at low energy and on a macroscopic scale, it is clearly more natural to
consider a gauge field description of gravity at higher energies.
The same conclusion would also raise the issue whether the metric tensor itself 
should be quantized as is done in conventional quantum gravity, with its
inherent problem of non-renormalizability, or that the gauge fields which lead
to the metric are to be quantized.  Our treatment here is purely classical
and does not attempt to address the deep quantization problem in this paper
\cite{8}. The cosmological constant is another issue that has to be explored. 
Although the present theory is identical to Einstein's gravity
in the classical sense, the quantum version of this theory might contain new
features. A gauge-invariant formulation of gravity and its coupling to matter
would allow new insights into the origin of the known gauge interactions.

\vspace{1.0in}
It is a pleasure to thank Professor Samuel W. MacDowell for stimulating
discussions.
\pagebreak[4]

\end{document}